\documentclass[sigconf]{acmart}
\acmSubmissionID{30}

\usepackage{algorithm}
\usepackage{algpseudocode}
\usepackage{gensymb}
\usepackage{graphicx}
\usepackage{subcaption}
\usepackage{xcolor}
\usepackage{amsmath}
\usepackage{amsfonts}
\usepackage{booktabs} 
\usepackage{xspace}
\usepackage{multirow}

\usepackage{tikz}
\AtBeginDocument{%
  }

\newcommand{\Deg}{\ensuremath{^\circ}\xspace}

\copyrightyear{2025}
\acmYear{2025}

\setcopyright{rightsretained}

\acmConference[MIG '25]{The 18th ACM SIGGRAPH Conference on Motion, Interaction, and Games}{December 3--5, 2025}{Zurich, Switzerland}

\acmBooktitle{The 18th ACM SIGGRAPH Conference on Motion, Interaction, and Games (MIG '25), December 3--5, 2025, Zurich, Switzerland}
\acmDOI{10.1145/3769047.3769058}
\acmISBN{979-8-4007-2236-3/2025/12}
\newenvironment{tightemize}{\vspace{-\topsep}\begin{itemize}\itemsep1pt \parskip0pt \parsep0pt}{\end{itemize}\vspace{-\topsep}}

\begin{document}

%%
%% The "title" command has an optional parameter,
%% allowing the author to define a "short title" to be used in page headers.
\title{Enhancing Foveated Rendering with Weighted Reservoir Sampling}

% Alternative Titles?

% Weighted Reservoir Sampling for Efficient Sample Accumulation for Foveated Rendering

% Foveated Rendering Enhancement using Weighted Reservoir Sampling

% Enhancing Foveated Rendering with Weighted Random Sampling

% Improving Foveated Rendering with Weighted Random Sampling

%%
%% The "author" command and its associated commands are used to define
%% the authors and their affiliations.
%% Of note is the shared affiliation of the first two authors, and the
%% "authornote" and "authornotemark" commands
%% used to denote shared contribution to the research.
\author{Ville Cantory}
\affiliation{%
  % \department{Department of Computer Science}
  \institution{University of Minnesota}
  \city{Minneapolis}
  \state{MN}
  \country{USA}}
\email{canto063@umn.edu}

\author{Darya Biparva}
\affiliation{%
% \department{Electrical and Computer Engineering}
  \institution{University of Minnesota}
  \city{Minneapolis}
  \state{MN}
  \country{USA}}
\email{bipar001@umn.edu}

\author{Haoyu Tan}
\affiliation{%
  \institution{University of Minnesota}
  \city{Minneapolis}
  \state{Minnesota}
  \country{USA}}
  \email{tan00213@umn.edu}

\author{Tongyu Nie}
\affiliation{%
  \institution{University of Minnesota}
  \city{Minneapolis}
  \state{Minnesota}
  \country{USA}}
  \email{nie00035@umn.edu}

\author{John Schroeder}
\affiliation{%
  \institution{University of Minnesota}
  \city{Minneapolis}
  \state{Minnesota}
  \country{USA}}
  \email{schr1294@umn.edu}

\author{Ruofei Du}
\affiliation{%
  \institution{Google}
  \city{San Francisco}
  \state{California}
  \country{USA}}
  \email{ruofei@google.com}

\author{Victoria Interrante}
\affiliation{%
  \institution{University of Minnesota}
  \city{Minneapolis}
  \state{Minnesota}
  \country{USA}}
  \email{interran@umn.edu}

\author{Piotr Didyk}
\affiliation{%
  \institution{Università della Svizzera italiana}
  \city{Lugano}
  \country{Switzerland}
}
\email{piotr.didyk@usi.ch}

%\author{Ruofei Du}
%\email{ruofei@google.com}
%\affiliation{%
%  \institution{Google}
%  \city{San Francisco}
%  \state{California}
%  \country{USA}
%}

%\author{Victoria Interrante}
%\email{interran@umn.edu}
%\affiliation{%
%  \institution{University of Minnesota}
%  \city{Minneapolis}
%  \state{Minnesota}
%  \country{USA}
%}

%\author{Piotr Didyk}
%\email{piotr.didyk@usi.ch}
%\affiliation{%
%  \institution{Università della Svizzera italiana (USI)}
%  \country{Switzerland}
%}%
\newif\ifCOMMENTS
% \COMMENTSfalse
\COMMENTStrue
\ifCOMMENTS
\newcommand{\yourname}[2]{\textcolor{blue}{#1\bf{ (yourname: #2)}}}
\newcommand{\ruofei}[1]{\textcolor{teal}{\bf{ (ruofei: #1)}}}

\else
\newcommand{\ruofei}[1]{}
\newcommand{\yourname}[2]{#1}
\fi
%%
%% By default, the full list of authors will be used in the page
%% headers. Often, this list is too long, and will overlap
%% other information printed in the page headers. This command allows
%% the author to define a more concise list
%% of authors' names for this purpose.
\renewcommand{\shortauthors}{Cantory et al.}

%%
%% The abstract is a short summary of the work to be presented in the
%% article.
\begin{abstract}

Spatiotemporal sensitivity to high frequency information declines with increased peripheral eccentricity.
Foveated rendering exploits this by decreasing the spatial resolution of rendered images in peripheral vision, reducing the rendering cost by omitting high frequency details.
As foveation levels increase, the rendering quality is reduced, and traditional foveated rendering systems tend not to preserve samples that were previously rendered at high spatial resolution in previous frames.
Additionally, prior research has shown that saccade landing positions are distributed around a target location rather than landing at a single point, and that even during fixations, eyes perform small microsaccades around a fixation point.
This creates an opportunity for sampling from temporally neighbouring frames with differing foveal locations to reduce the required rendered size of the foveal region while achieving a higher perceived image quality.
We further observe that the temporal presentation of pixels frame-to-frame can be viewed as a data stream, presenting a random sampling problem.
Following this intuition, we propose a Weighted Reservoir Sampling technique to efficiently maintain a reservoir of the perceptually relevant high quality pixel samples from previous frames and incorporate them into the computation of the current frame.
This allows the renderer to render a smaller region of foveal pixels per frame by temporally reusing pixel samples that are still relevant to reconstruct a higher perceived image quality, while allowing for higher levels of foveation.
Our method operates on the output of foveated rendering, and runs in under 1\,ms at 4K resolution, making it highly efficient and integrable with real-time VR and AR foveated rendering systems.

% Start with bullet items
% Introduce the essence of our insight
% Make sure main contributions are clearly stated and shown in here
% Say what we do in the paper, like a summary of our primary contributions at the end of Intro

% Vicki's notes: follow the email she'll send so that we can make sure the abstract is pretty bulletproof . Lead with the problem, then explain how WRS is the solution.
%

%\ruofei{}{I just realized that "Persistence of vision" may be one motivation of the work..lots of text on motivation -> could be more concise, siggraph reviewers understands foveated rendering; need more words in method novelty and how this is achieved in real-time. One sentence summary is missing.}

% This is a good sentence to put somewhere else
%Weighted reservoir sampling from data streams is a weighted random sampling method that selects a subset from a stream of weighted items, where each item's probability of being sampled is proportional to its weight.

\end{abstract}

%%
%% The code below is generated by the tool at http://dl.acm.org/ccs.cfm.
%% Please copy and paste the code instead of the example below.
%%



% \ccsdesc[300]{Do Not Use This Code~Generate the Correct Terms for Your Paper}
% \ccsdesc{Do Not Use This Code~Generate the Correct Terms for Your Paper}
% \ccsdesc[100]{Do Not Use This Code~Generate the Correct Terms for Your Paper}

%%
%% Keywords. The author(s) should pick words that accurately describe
%% the work being presented. Separate the keywords with commas.
%% A "teaser" image appears between the author and affiliation
%% information and the body of the document, and typically spans the
%% page.
\begin{teaserfigure}
\centering
    \includegraphics[width=\columnwidth]{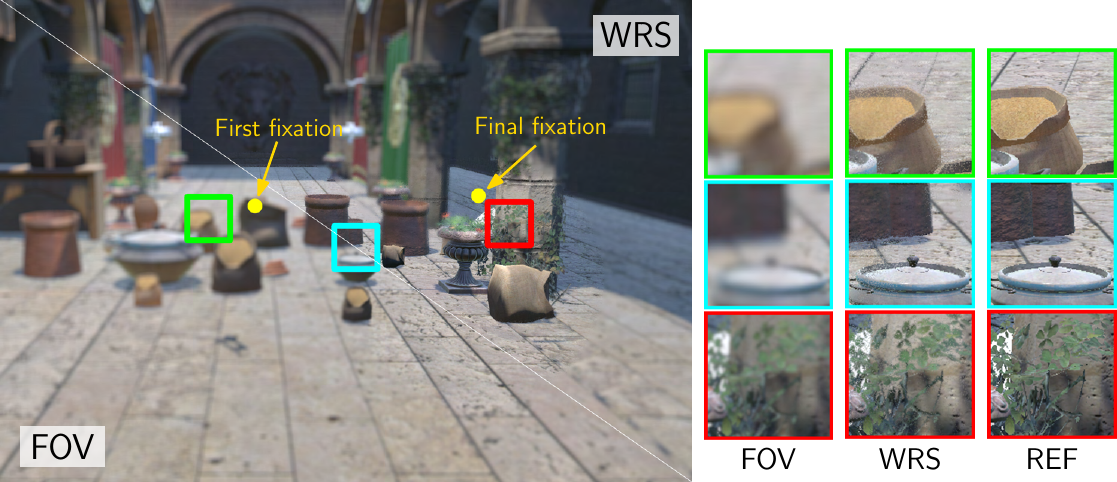}
    \caption{ 
        Our method temporally enhances the output of foveated rendering.
        Given an eye-tracked scanpath, traditional rasterization-based foveated rendering (FOV) methods (left) do not preserve previously rendered foveal samples.
        Our enhancement method (right) uses Weighted Reservoir Sampling (WRS) to stochastically preserve high-quality rasterized samples across time, leading to higher overall image quality.
        The images above feature a 5\Deg fovea and were designed to be viewed at 70\Deg field-of-view.
        Sponza courtesy of Casual Effects \cite{McGuire2017Data}.
    }
        \label{fig:teaser_figure}
\end{teaserfigure}

% FUN IDEA: Could we try near the teaser or in some of the other sections showing off a big blowup image with scenes and their FovVideoVDP maps and CGVQ1M maps?

% \received{20 February 2007}
% \received[revised]{12 March 2009}
% \received[accepted]{5 June 2009}

%%
%% This command processes the author and affiliation and title
%% information and builds the first part of the formatted document.
\maketitle
% \hypersetup{
%     colorlinks=true,
%     linkcolor=blue,
%     filecolor=magenta,      
%     urlcolor=cyan,
%     pdfpagemode=FullScreen,
%     }

% For arxiv only:
\noindent\textcopyright~2025 Copyright held by the owner/author(s). This is the author's version of the work. It is posted here for your personal use. Not for redistribution. The definitive version of record was published in \textit{The 18th ACM SIGGRAPH Conference on Motion, Interaction, and Games (MIG '25)}, \url{https://doi.org/10.1145/3769047.3769058}.

% \urlstyle{same}
\section{Introduction}
% \begin{itemize}
%     \item Introduce HVS, and how it allows for FR
%     \item oveated rendering can hinder perceived image quality, especially temporally.
% Mention mitigation methods (Noise2022 for instance, the raytracing/path tracing ones too)
% \item We look at this from a probabilistic point of view using reservoir sampling of weighted streams
% \item Say structure of paper
% \end{itemize}

Real-time rendering requirements continue to grow as shading model accuracy and spatiotemporal display standards increase.
The top of the line consumer VR headset, the Varjo XR-4\footnote{Varjo XR-4: \url{https://varjo.com/products/xr-4/}} boasts a resolution approximately 3.5× what a traditional 4K (3840 × 2160) resolution displays.
Efficient rendering for these displays remains challenging, necessitating the development of ways to lower computational costs, ideally without an impact on perceptual quality.

Foveal vision is an approximately 2\Deg region of vision with high visual acuity, with the rest of the visual field suffering from lower visual acuity \cite{bergmann1854kenntniss, thibos1996characterization}.
To save computational resources, foveated rendering takes advantage of the reduction in visual acuity by rendering lower quality samples in regions that diverge from the fovea.
This is commonly done by reducing the sampling rate in peripheral vision, eliminating high-spatial frequency details in peripheral portions of the scene; yet it has been consistently observed that the amount of foveation tenable depends on the scene viewed, with some modulators for this including the frequency content of the image viewed \cite{patney2016towards}, luminance \cite{tursun2019luminance}, attention subtended to different visual fields \cite{krajancich2023towards}, and more.
It has also been shown that the presence of high frequency details, even in peripheral vision, can be detected and are important for visual quality, with some works opting to reconstruct image metamers -- high frequency details that match the statistics of the scene content, but do not need to contain the same content as the original ground truth in order to have a high perceptual quality \cite{tariq2022noise, walton2021beyond}.

Most current approaches fail to persist previously rendered high-resolution samples, replacing them with lower resolution samples as gaze location shifts.
If a fragment were previously rendered at a high spatial resolution, and visibility is maintained but the fragment is not otherwise changed due to occlusions or shading changes, replacing it with a lower quality version is unnecessary.
Discarding recently rendered high spatial resolution samples also overlooks common oculomotor behaviour; Saccades rarely land precisely on their intended target \cite{van2020transsaccadic}, and are corrected with microsaccades and ocular drift.
Further, fixations themselves are interspersed with microsaccades rather than truly fixating at a single location \cite{zuber1965microsaccades, van2007sources}.
The Human Visual System (HVS) integrates information temporally, meaning transient high-resolution details can contribute to the final image perception.
Thus, discarding high-quality samples immediately upon gaze shifts may discard perceptually relevant details.

Naturally, this creates an analogue to a data streaming problem, where perceptually important pixel samples arrive over many frames and must be selectively retained.
Reservoir sampling is a family of sampling algorithms designed to maintain a representative subset of elements from a stream without preprocessing the stream \cite{fan1962development}, and was recently introduced to computer graphics for sampling light paths for real-time-raytracing (ReSTIR) \cite{bitterli2020spatiotemporal}.

% By utilizing a variation of reservoir sampling, \textit{Weighted Reservoir Sampling} (WRS), where each stream element has a weight $w$ drawn from a distribution, $p$, describing its importance, we show how we temporally and efficiently accumulate higher quality samples using WRS from a data stream.
% Building from ideas introduced by ReSTIR, we propose a WRS strategy tailoured to eye-tracked foveated rasterization, and maintain a reservoir of well-sampled pixels from recently rendered frames, prioritized by those we infer to be 
%  most perceptual relevant.
% Unlike ReSTIR, we focus on foveated rasterization rather than trying to get a uniformly high quality sampling across all pixels.

Similar to ReSTIR, we employ \textit{Weighted Reservoir Sampling} (WRS), a variant of reservoir sampling in which each stream element is assigned a weight $w$ drawn from a distribution $p$ representing its importance.
However, we focus on foveated rasterization and temporally accumulating higher quality samples from a foveated data stream, prioritizing samples by perceptual relevance.

We introduce a novel temporal bias function to bias the reservoir based on our interpretation of the reservoir sampling process as a Bernoulli Trial to prevent stale samples from inappropriately persisting in the reservoir.
Our method significantly reduces the number of foveal pixels that have to be rendered per-frame while maintaining high perceptual quality.

Our primary contributions are:
\begin{tightemize}
    \item Introducing Weighted Reservoir Sampling (WRS) to foveated rasterization to temporally accumulate high-quality pixel samples; 
    \item  Proposing a new temporal bias function based on Bernoulli Trials to reweight reservoir samples without dependence on absolute or relative time that samples were selected for inclusion in the reservoir;
    \item Validation of our approach via user studies to improve foveated text reading speed, and improved image quality over foveated rendering and foveated rendering with Temporal Antialiasing (TAA).
\end{tightemize}

\section{Related Work} 
\label{sec:related_work}
Our work is inspired by a wide body of prior research in foveal and peripheral vision and foveated rendering.
In this section, we introduce prior perceptual work about foveal and peripheral vision in \S~\ref{subsec:perceptual_bkg}.
We proceed to discuss a body of previous work in foveated rendering in \S~\ref{subsec:foveated_rendering}, and prior temporally adaptive rendering work in \S~\ref{subsec:shading_reuse_bkg}

\subsection{Perceptual Background} \label{subsec:perceptual_bkg}
% Include stuff about https://dl.acm.org/doi/pdf/10.1145/3641519.3657420 and saccades within a foveal / fixation point. 
% We are not constantly focusing. 
% Include stuff about eye movements (we have many many many saccades per second - show this off!). 
% include some stuff about loss of visual acuity in this section, and in FR, talk about the tech but keep loss of details here

\subsubsection{Foveal and Peripheral Vision} 

The Human Visual System (HVS) features a 2\Deg \space  central region of vision capable of high visual acuity, with the rest of the visual field perceiving progressively lower visual acuity \cite{strasburger2018ewald, aubert1857beitrage, lettvin1976seeing}.
The physiological reason for this is there exists a region of the eye with a dense field of photoreceptors (cones), deemed the \textit{fovea centralis} \cite{bergmann1854kenntniss, bergmann1857anatomisches}; the rest of the eye is less densely populated with photoreceptors (mostly rods) resembling a blue noise distribution \cite{curcio1990topography, curcio1990human, lanaro2020blue}.
While there is a rapid optical decrease in visual acuity and retinal ganglion-cell density as eccentricity from the fovea increases \cite{watson2014formula}, neural factors degrade visual acuity more rapidly \cite{navarro1993modulation, williams1996off}.
Moreover, this can lead to a range of spatial frequencies in peripheral vision for which certain frequencies are detectable, but low-level details are not resolvable \cite{thibos1987vision, sere2000nonhomogeneous}.

\subsubsection{Saccades}

Saccades are large eye movements \cite{javal1878essai} under which visual sensitivity is reduced \cite{erdmann1898psychologische, richards1969saccadic}, producing retinal blur during eye movements  \cite{volkmann1962vision, dodge1900visual}.
Saccades typically occur 2-4 times per second \cite{otero2013oculomotor} and last under 50ms \cite{robinson1964mechanics}.
%It has also been observed that if eye gaze is at rest and the environment is moved rapidly, such as with a mirror, suppression effects similar to saccadic eye movements can be observed \cite{woodworth1906vision, woodworth1939}.
While a saccade can subtend up to 200\Deg \cite{robinson1964mechanics}, a \textit{microsaccade} often moves around 1\Deg of visual angle during fixation \cite{zuber1965microsaccades}.
Microsaccades can enhance the perception of fine details during fixation \cite{rucci2007miniature};
%By realigning gaze towards the fixation target, they counteract eye drift so visual lines are kept in the center of the fovea \cite{moller2006contribution}.
after a saccade from one fixation point to another, many small ocular drifts and microsaccades can involuntarily follow \cite{rucci2015control}.
%Interestingly, even microsaccades can introduce saccadic suppression \cite{zuber1964saccadic}. -- I don't want to include this because I'll need to find another reference or many characterizing microsaccades  and how they might or might notintroduce suppression. This section already got longer than I meant 
Visual acuity is not just reduced during the saccade, but for 50-500ms after saccadic landing (completion of the saccade) depending on the spatial frequency of the content \cite{kwak2024saccade}.

In our work, we seek to exploit microsaccades by utilizing them to expand the effective fovea size displayed to the user.

\subsection{Foveated Rendering} \label{subsec:foveated_rendering}
Foveated rendering achieves computational speedups by exploiting the reduction in peripheral visual acuity.
Early works on foveated rendering focused on using image and video compression to improve bandwidth \cite{kortum1996implementation, tsumura1996image}.
Gaze-contingent mesh simplification methods have also been thoroughly explored \cite{ohshima1996gaze, surace2023gaze}; however, as fragment shading tends to be much more computationally expensive than geometry processing \cite{he2014extending}, most efforts seeking performance benefits focus on reducing shading  costs.

Eccentricity-dependent shading rate reductions, driven by the sharp drop in visual acuity as retinal eccentricity increases, has been the focus of much modern foveated rendering research.
Commonly used thresholds for the detection of spatial frequency reductions (accomplished with reductions in fragment shading rates) were first formalized in \citet{guenter2012foveated}, and further improved in subsequent works \cite{vaidyanathan2014coarse, patney2016towards}.
Similarly, \citet{meng2018kernel} achieved higher performance benefits by utilizing a 2-pass foveated rendering pipeline specifically built for deferred shading.
As the amount of foveation tolerable depends on many factors such as eye motion, contrast, luminance, image content, attentional load, and more \cite{albert2017latency}, there have been some explorations into finding better adaptive sampling patterns to drive a reduction in peripheral shading rates based on these factors \cite{stengel2016adaptive, tursun2019luminance}, resulting in sample count reductions in raytracing and path tracing \cite{weier2016foveated, swafford2016user, polychronakis2021emulating} and optimizations in split-rendering systems \cite{cantory2023image}.

Recent works have added high frequency details to peripheral regions to better preserve the perceptual experience of full-resolution rendering under foveation.
\citet{walton2021beyond} constructed foveated image metamers - images that are different from the ground truth, but will be perceived as the same - and \citet{tariq2024towards} explored motion metamerism. 
\citet{tariq2022noise} incorporated adding back in Gabor noise to achieve higher levels of foveation, offset by the noise metamer.
%\citet{tariq2024towards} synthesized motion energy to construct motion metamers for foveated rendering.
DeepFovea \cite{kaplanyan2019deepfovea} utilized Generative Adversarial Networks \cite{goodfellow2014generative} to reconstruct foveated images with extremely sparse sampling.
For a comprehensive review of foveated rendering, we refer readers to \cite{wang2023foveated}.

Outside of raytracing and pathtracing, we are unaware of any prior gaze-contingent rendering works stochastically accumulating well-sampled pixels to temporally increase image quality.
Within a rasterization context, it stands to reason that if only the fixation point and not the viewport have drastically changed frame to frame, pixels that were previously rendered at a high shading rate can be reused to expand the number of well-sampled pixels without unduly increasing the cost to render them.
In our work, we utilize a Reservoir Sampling algorithm to achieve higher quality temporal image reconstruction from a stream of foveated images.

\subsection{Temporally Adaptive Rendering}
\label{subsec:shading_reuse_bkg}
% Stuff for temporal reuse in 1990s, TAA, TAAU, DLSS, 
Temporal reuse to achieve higher image quality has been widely used in computer graphics applications.
Intuitively, most applications change very little in the 8-16 ms that occurs between frames, meaning that much of the information computed from previous frames can still be relevant to the current one, achieving possibly an increased framerate by reducing the rendering cost, or improving the image quality of each frame.

%Image warping using forward projection has been used in raycasting to warp the previous frame's output into the current frame, with some additional rendering required to solve for disocclusions \cite{qu2000image}.
%Raytraced animiation sequences have exploited temporal coherence to produce high-quality yet efficient motion blurring, and to reduce the cost of global illumination by adding each sample point's colour contribution to relevant pixels by analyzing camera and scene motion \cite{havran2003efficient}.

The Reverse Reprojection Cache (RRC) proposed by \citet{nehab2007accelerating} and \citet{scherzer2007pixel} for reprojection with shadow maps, describes a framebuffer that contains pixel data visible at a given frame, and the scene depth in screen space rendered from the camera's point of view.
RRC's can be used to determine the location where a pixel was in the previous frame, as well as the visibility.
However, non-linear warping often incurs cache misses due to visibility changes or view-dependent shading changes.
Re-rendering or inpainting can occur \cite{marroquim2007efficient}, but cheaper methods of shading reuse have been explored using a \textit{compact geometry buffer} that stores and reuses shading samples separate from visibility, \cite{liktor2012decoupled}, or through methods that adaptively change the shading rate using error estimation based on spatial frequency analysis \cite{yang2019visually}.

A common usage of temporal reuse in consumer applications has been Temporal Antialiasing (TAA).
Originally introduced by \citet{korein1983temporal} to reduce temporal aliasing, modern TAA methods often seek to reduce spatial aliasing by using temporal samples to achieve supersampling.
TAA reuses subpixel samples accumulated from previous frames to achieve supersampling, generally relying on jittered subsamples to be generated each frame.
This leads to much better integration over a pixel's domain, and the samples are summed using a weighted average and stored in a history buffer representing the samples that have been rendered.
Reverse reprojection is used to map samples to their previous location; however, when the history buffer is invalidated by something like large shading changes or occlusions / disocclusions, the history buffer may be invalidated at an affected pixel, resetting the effective sample count.

Temporal Upscaling (TAAU) seeks to reduce the sample rate per pixel to under one per pixel for the target resolution.
It accumulates lower resolution shading results to produce a higher resolution over many frames. 
However, as noted by \citet{yang2020survey}, ghosting and artifacts can occur in TAA solutions if the history buffer is not carefully rectified when it may become invalidated.
More recently, DLSS has used deep neural networks to achieve fast supersampling, relying on reverse reprojection techniques and reusing the previous frame to supersample the next frame.
For a comprehensive overview of temporal reuse, we refer readers to \citet{scherzer2011survey, yang2020survey}.

\section{Mathematical Preliminaries} \label{sec:preliminaries}
% One key thing about ReSTIR is that every frame, they see a stream of candidate rays. For them, it's useful to know what will be the last sample included in the reservoire as the chosen candidate ray. However, for us / in rasterization, we are only seeing a 1 element stream per pixel per frame. Our stream is temporal, frame-to-frame, so we never have any interest in a "last selected item" in the reservoir.

This section outlines the mathematical basis of our reservoir sampling method and highlights prior applications of weighted reservoir sampling in computer graphics.

\subsection{Weighted Reservoir Sampling} \label{subsec:wrs}
% Move some stuff from the previously background section into here. But keep it brief.
Reservoir sampling is a technique to efficiently sample $n$ items from a population of potentially unknown size $N$.
%Reservoir sampling has been shown to be efficient for sampling in a single pass over the data stream \cite{fan1962development}, and requires only constant memory.
%It only requires constant memory and a single pass over data.
Originally introduced by \citet{fan1962development} for sequential sampling from tape drives in a single pass, reservoir sampling was later formalized for uniform sampling from data streams \cite{vitter1984faster, vitter1985random}.
It has also been generalized to allow non-uniform sampling probabilities, leading to Weighted Reservoir Sampling (WRS) \cite{chao1982general}.
This variation is particularly well-suited to cases where evaluating large sets of candidate samples is expensive and real-time performance is required. 
It can also be applied to sampling problems where the input data is evolving \cite{aggarwal2006biased, efraimidis2015weighted}.
These characteristics make WRS an attractive choice for computer graphics applications, where large and expensive sets of candidate samples, such as light paths or pixel contributions, must be considered under tight computational budgets.
In such cases, WRS enables efficient sampling from a dynamically evolving stream in a single pass, avoiding repeated evaluations of the full set.
In particular, WRS enables sampling from a probability density function, $p$, where each candidate's weight reflects its relative importance. 
Because $p$ is often expensive to exactly evaluate in most rendering contexts, samples are typically drawn from an approximation, $\hat{p}$ \cite{bitterli2020spatiotemporal, ouyang2021restir, wyman2021rearchitecting}.

\newcommand{\Exp}[1]{\exp\left(#1\right)}
\newcommand{\sample}{\boldsymbol{x}}
\newcommand{\wsum}{w_{\textrm{sum}}}
\newcommand{\rand}{\textit{rand}}

\newcommand{\updater}{\textit{updater}\xspace}
\newcommand{\inst}{\texttt{inst}\xspace}
\newcommand{\dataset}{\mathcal{D}}
\newcommand{\workflow}{\mathcal{W}}
\newcommand{\triplet}{(q, \workflow, r)}
\newcommand{\ul}[1]{\underline{#1}}

Formally, given a stream of candidate samples $\sample_{1}, \dots, \sample_{N}$, each associated with a positive weight $w_{i} \propto \hat{p}(x_i)$,
%, where $\hat{p}$ is an approximation of the true importance distribution $p$.
% Each item $x_{i}$, for $i= 1,..., N$ is associated with a strictly positive weight $w_{i}$ which represents the relative importance drawn from $\hat{p}$.
the goal of WRS is to maintain a representative sample of size $n$ from this stream.
For the remainder of this paper, we focus on cases where the reservoir operates under tight memory constraints, and thus we maintain a reservoir containing only a single sample ($n = 1$) and that the total length of the data sequence, $N$, is unknown.
The items of the population are not assumed to be unique, but they are distinguishable.
%WRS assumes that the full set of items and their weights is either unavailable or expensive to evaluate upfront.

% \textcolor{blue}{Weights in WRS fall into at least two common interpretations: In the first interpretation, an items' weight determines the probability that the item is in the final reservoir.
% In this situation, the stream evolution is not of concern, only the final $n$ samples in the reservoir are relevant after viewing all $N$ items.
% In the second interpretation, each items' weight determines the selection probability at the time the item is evaluated.
% This does not, however, mean that the sample will be in the \textit{final} reservoir after all items have been processed \cite{efraimidis2015weighted}.}

The core algorithms for reservoir sampling are A-Chao \cite{chao1982general} and A-ES \cite{efraimidis2006weighted, efraimidis2015weighted}. 
% IMPORTANT: A-Chao is Interpretation 1, A-ES is 2.
In A-Chao, when a new sample $\sample_i$ is encountered, its selection probability is computed and used to decide whether it replaces the current reservoir entry, as:
% In A-Chao, each item $x_i$ with a corresponding weight $w_i$ has an inclusion probability proportional to the currently evaluated items' weight compared to the weighted sum of all the samples the reservoir has processed.
% This simple formulation is shown in Equation \ref{eq:chao-selection-probability}
\begin{equation}
    p(\text{select }   \sample_i) = \frac{w_i}{\sum_{j=1}^{i} w_j}.
    \label{eq:chao-selection-probability}
\end{equation}
% We present a modified presentation of A-chao for reservoirs of size $n=1$ in Algorithm~~\ref{alg:combine}.
% Thus, a modified presentation of A-Chao is presented in Algorithm~\ref{alg:combine}.
% In this presentation, it is assumed that the reservoir has already been initialized with $n$ candidates, and that there exists a reservoir structure that samples $n$ candidates that stores $x, w, w_{sum}$ (the weighted sum of all evaluated candidates), and $m$, the number of total evaluated candidates.
The key insight is that if reservoir items are sampled proportionally to their weights, then uniformly replacing one preserves this distribution.
We present a modified version of A-Chao for a reservoir of size $n = 1$ in Algorithm~\ref{alg:update}.
%where the update depends on comparing to a random float $rand \in [0, 1]$. 

\begin{algorithm}
\caption{A-Chao Update Reservoir} \label{alg:update}
\begin{algorithmic}[1] 
\Function{update}{$x_{i}$, $w_{i}$}
    \State $\wsum \gets \wsum + w_{i}$
    \If{$\rand \cdot \wsum < w_{i}$}
        \State $\sample \gets \sample_{i}$
        \State $w \gets w_{i}$
    \EndIf
\EndFunction
\end{algorithmic}
\end{algorithm}
In this formulation, it is assumed that a reservoir structure maintains three quantities: the current sample in the reservoir $x$, its weight $w$, and the cumulative weight sum $w_{sum}$ of all observed candidates.
$x_{i}$ is the sample that is being evaluated and $\rand \sim \mathcal{U}[0, 1] $.
% This allows for the sampling decision to be made with the only memory requirements for the reservoir being the sample itself, its weight, and the weighted sum of all previous samples.
This allows the sampling decision to be made with minimal memory requirements.
%This allows for the sampling decision to be made with the only memory requirements for the reservoir being $\sample$, $w$, and $\wsum$.
% This obeys the first common interpretation of weights for reservoir sampling algorithms.

% For applications where weights are interpreted as importance scores rather than the final inclusion probabilities, \citet{efraimidis2006weighted} and \citet{efraimidis2015weighted} introduced Algorithm A-ES.
Algorithm A-ES introduced by \citet{efraimidis2006weighted} takes a different approach.
% For applications where weights are interpreted as importance scores rather than the final inclusion probabilities, \citet{efraimidis2006weighted} introduced Algorithm A-ES.
They proposed a sampling schema dependent on a random key, $k_i = \rand_{i}^{1 / w_{i}}$ for stochastically updating a reservoir item.
% , where $u_i$ is an independent random variable drawn from a uniform distribution in $[0, 1]$.
The reservoir, at each step, will contain $n$ largest keys from a weighted random stream.
Larger weights produce larger denominators in the exponent.
They argue that a uniform random variable can be ``amplified'' by raising it to an appropriate power.
% A version of this algorithm for $n=1$ is given in Algorithm~\ref{alg:Aes}.
% \begin{algorithm}
% \caption{A-ES Update Reservoir}
% \begin{algorithmic}\label{alg:Aes}
% \State \textbf{function} update($\sample_{i}$, $w_{i}$)
%     \State $k_{i} \gets \rand_{i}^{{1 / w_{i}}}$
%     \If{$k_{i} >k$}
%         \State $\sample \gets \sample_{i}$
%         \State $k \gets k_{i}$
%     \EndIf
% \end{algorithmic}
% \end{algorithm}
In this formulation, it is assumed that reservoir structure maintains the sample $x$ and the key for this sample $k$, but the weighted sum of all previously evaluated candidates is not stored, making it more difficult to temporally evaluate the sample quality.

The aforementioned sampling algorithms focus on unbiased sampling; yet to support cases of reservoir sampling from an evolving data stream, a temporal bias can be applied either into the weight of items currently in the reservoir, or into the weight of newly evaluated items \cite{efraimidis2015weighted}. 
A commonly used bias function is the \textit{memory-less} temporal bias  introduced by \citet{aggarwal2006biased}: 
% \citet{aggarwal2006biased} introduces a \textit{memory-less} temporal bias function (Equation~\ref{eq:memory-less}) that assumes reservoir sampling is done from a uniformly distributed set of random variables, where all $w_{i}$ are equal;
%\citet{aggarwal2006biased} introduces a bias assuming that reservoir sampling is done from a uniformly distributed random variables set, where all $w_i$ are equal. 
%A very popular class of temporal bias functions are \textit{memory-less} bias functions, with one such case defined in Equation~\ref{eq:memory-less}. 
% \begin{equation}
%     f(i, j) = e^{-\lambda (j - i)}
%     \label{eq:memory-less}
% \end{equation}
% This equation represents an exponential \textit{damping function} \cite{diop2024rps}, where $\lambda$ is a predefined bias rate, $i$ is the index of the sample currently in the reservoir, and $j$ is the index of the sample currently being evaluated against $x_i$.
% %When sampling from uniformly distributed random variables, where all $w_i$ are equal, the probability that $x_i$ is included in the reservoir is simply $\frac{n}{i + 1}$.
% Given a bias function $f(i, j)$, samples in the reservoir ($w$) can be reweighted such that more recently evaluated samples are more likely to be selected for inclusion.
% Memory-less bias functions reduce the probability of each reservoir sample to persist in the reservoir as relative time passes, but the absolute time of insertion / comparison do not matter.
\begin{equation}
    f(j, i) = e^{-\lambda (i - j)}.
    \label{eq:memory-less}
\end{equation}
It assumes uniform sampling with equal weights $w_i$ and represents an exponential \textit{damping function} \cite{diop2024rps}, where $\lambda$ is a predefined bias rate, $j$ is the index of the sample currently in the reservoir ($x$), and $i$ is the index of the sample currently being evaluated against $x$.
%When sampling from uniformly distributed random variables, where all $w_i$ are equal, the probability that $x_i$ is included in the reservoir is simply $\frac{n}{i + 1}$.
Given a bias function $f(j, i)$, samples in the reservoir can be reweighted such that more recently evaluated samples are more likely to be selected for inclusion: $w = w \cdot f(j, i)$. 
Memory-less bias functions reduce the probability of each reservoir sample to persist in the reservoir as relative time passes, but the absolute time of insertion / comparison does not matter.

\subsection{WRS in Computer Graphics}\label{sec:wrs_graphics}
%Reservoir sampling algorithms have commonly appeared for query optimization and data mining.
WRS combined with resampled importance sampling (ReSTIR) has been extensively used to evaluate real-time raytracing ray queries \cite{bitterli2020spatiotemporal, wyman2023gentle}.
ReSTIR spatiotemporally resamples candidate light samples to progressively increase the sample quality that leads to faster and higher quality real-time raytracing.
Their version of WRS uses replacement, meaning that the population of candidates to be evaluated is unaltered after an item is selected.
This version of reservoir sampling is usually done with Algorithm A-Chao \cite{meligrana2024investigating, efraimidis2015weighted}.

% Their weight interpretation based on A-Chao falls into the first interpretation described in Section~\ref{subseq:wrs}

They contend that WRS enables efficient spatiotemporal reuse by allowing for spatiotemporally neighbouring reservoirs to be combined without needing to reevaluate each sample distribution.
Since a reservoir encodes its selected sample, sample weight, and the cumulative sum of all evaluated candidates, combining reservoirs reduces to treating each reservoir as a single weighted sample.
For instance, combining reservoir $R_{1}$ that has seen $m_{1}$ candidates with neighbour $R_{2}$ that has seen $m_{2}$ candidates is equivalent to performing WRS on a stream of two samples.
This preserves the invariant of WRS and increases the effective sample count to $m_{1} + m_{2}$.
When combining $h$ reservoirs, the sample count increases to $m_{1} + \cdots + m_{h}$.

We show the Combine Reservoirs algorithm introduced by ReSTIR \cite{bitterli2020spatiotemporal} in Algorithm~\ref{alg:combine} without resampling against differing target distributions.
As it is most common to compare only two reservoirs at a time (i.e., compare a current pixel's reservoir to its temporal neighbour or to a spatial neighbour), the presented algorithm is shown in simplified form.
\begin{algorithm}
\caption{Combine Reservoirs}\label{alg:combine}
\begin{algorithmic}[1] % [1] adds line numbers
\Function{combine}{$R_{1}$, $R_{2}$}
    \State $R_{\text{out}} \gets \begin{cases}
        R_1 & \text{if } (\rand \cdot (R_{1_{\wsum}} + R_{2_{\wsum}}) \le R_{1_{\wsum}}) \\
        R_2 & \text{otherwise}
    \end{cases}$
    \State $R_{\text{out}_{\wsum}} \gets R_{1_{\wsum}} + R_{2_{\wsum}}$
    \State $R_{\text{out}_{m}} \gets R_{1_{m}} + R_{2_{m}}$
    \State \textbf{return} $R_{\text{out}}$ % Assuming you want to return the combined reservoir
\EndFunction
\end{algorithmic}
\end{algorithm}

\section{Method Overview} \label{sec:method}
% The implication within a rendering context is that if a sample has lived in the reservoir for a long period of time, it may not be perceptually representative 

%% Make it VERY clear here that  we are utilizing things from ReSTIR, and show what we've changed to solve our NEW problem of rasterization rendering. But we don't want to invite the reader to compare us to restir and ask us why not just use restir for this. We're not improving on restir, we're solving a different problem.

Our main contribution is to introduce foveated rasterization as a reservoir sampling problem with temporal sample accumulation, enabling us to reduce the rendered size of the fovea per generated frame while temporally increasing the number of high-quality pixels displayed to the user.
While there exist reservoir sampling approaches for real-time raytracing \cite{bitterli2020spatiotemporal}, rasterization poses different problems: We are not sampling potential light paths and trying to select the best one to trace.
Instead, each frame, we are given a rasterized pixel that has a given level of foveation applied to it, and the problem becomes whether to include that pixel in the final output displayed to a user.
With an input of a sequence of foveated images, we demonstrate how the reservoir allows us to adaptively accumulate high-quality temporal samples. 

% When a sequence of foveated images are seen, our method will temporally process each image through the reservoir to accumulate higher quality samples while maintaining temporal stability.
% As eye gaze or the viewport moves, different pixels will be rendered as high quality samples (ie, at a high fragment shading rate, or without Gaussian blur depending on implementation), causing a change in the rendering quality per pixel frame-to-frame.

\subsection{WRS for Foveated Rendering} \label{sec:wrsgraphics}
For each pixel location, the sequence of incoming values across frames forms a data stream, where each frame contributes a new candidate sample. 
The weight $w_{i}$ associated with each candidate reflects its spatial proximity to the current gaze location, prioritizing samples that were rendered with greater visual fidelity this frame. 
Our method applies WRS independently at each pixel location, treating each as a separate stream. 
The reservoir retains an estimate of the most representative rendered sample observed so far for that pixel location, probabilistically determined by the weight of each candidate compared to the weighted sum of all previously evaluated candidates. 
Consequently, each frame consists of temporally accumulated samples that preserve perceptual relevance.

\begin{figure}[htbp]
    \centering
    \includegraphics[width=\columnwidth]{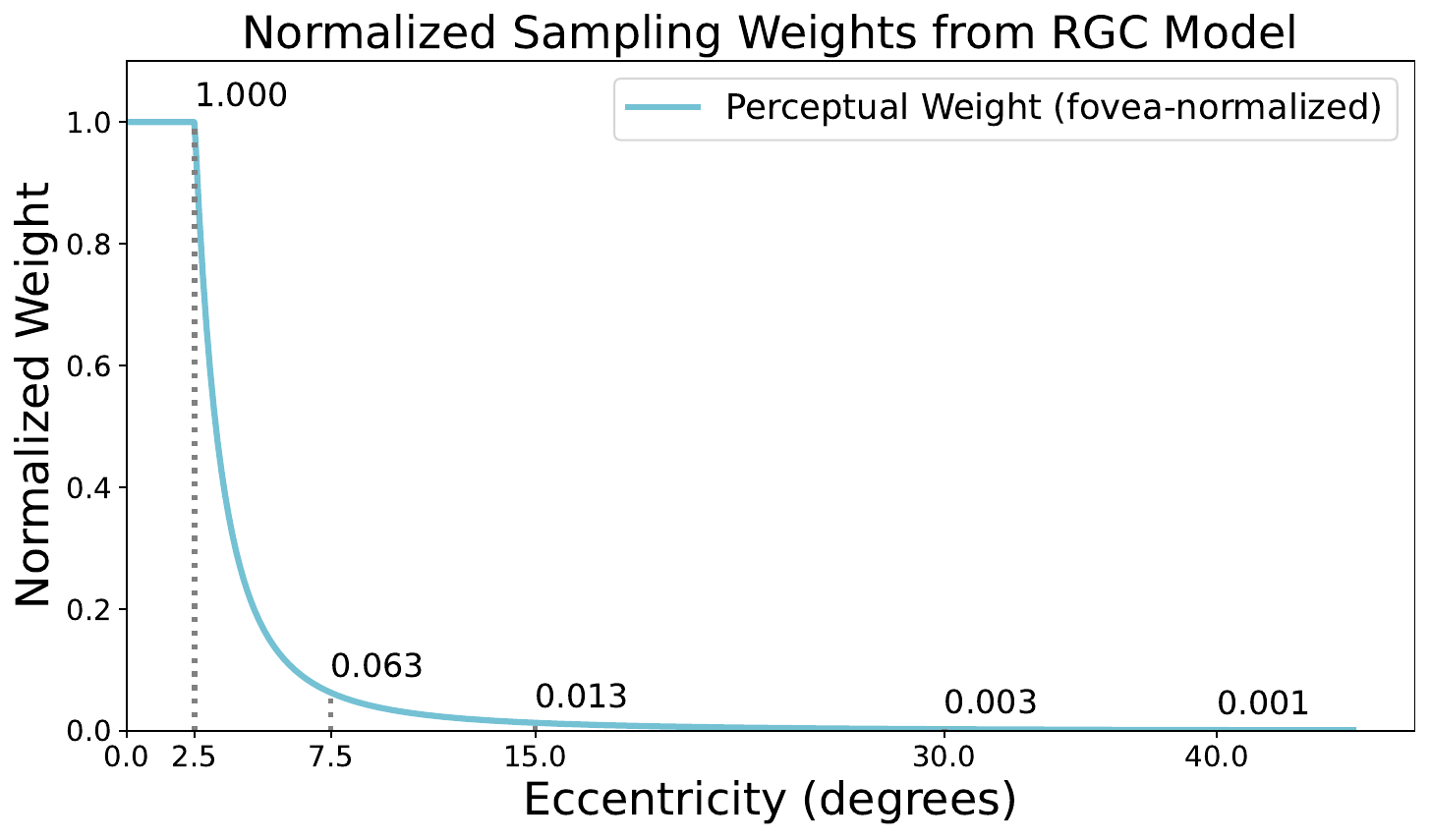}
    \caption{
    Visualization of Equation~\eqref{eq:phat} weights. The distribution is drawn from an approximation of visual acuity falloff given by \citet{watson2014formula}, normalized by the peak retinal ganglion cell density at a 2.5\Deg  eccentricity from the fovea.
    }
    \label{fig:phat_figure}
\end{figure}

A proposed distribution $\hat{p}(x)$ from which weights are drawn should be efficient to evaluate and sample from while providing an approximation of the target distribution $p(x)$.
Visual acuity thresholds for a given stimulus depend on a variety of factors such as luminance, spatial frequency of the stimuli, contrast, and more.
Several thorough contrast sensitivity functions (CSFs) provide models to approximate this \cite{krajancich2021perceptual, mantiuk2022stelacsf, cai2024elatcsf, ashraf2024castlecsf}, but computing the parameters to evaluate them becomes more expensive.
Instead, we derive our $\hat{p}$ by sampling from the distribution of retinal cone density modeled by \citet{watson2014formula} on data from \cite{curcio1990human, curcio1990topography} given in Equation~\eqref{eq:cone_retinal_density}.
\begin{align}
d(e) ={}&\ 
2 d_c(0) \left(1 + \frac{e}{r_m} \right)^{-1} \nonumber \\
&\cdot \left[
a_k \left(1 + \frac{e}{r_{2,k}} \right)^{-2}
+ (1 - a_k) \exp\left(-\frac{e}{r_{e,k}}\right)
\right]
\label{eq:cone_retinal_density}
\end{align}
We use the parameters reported by \citet{watson2014formula}: $a_k = 0.9851, r_{2,k} = 1.058, r_{e,k} = 22.14, d_{c}(0) = 14804.6$, and $e$ denotes the eccentricity.
We then normalize Equation~\eqref{eq:cone_retinal_density} by the peak retinal cone density at the edge of our fovea bounds if outside the fovea, or assign a weight of $1$ if inside, producing Equation~\eqref{eq:phat}, and visualized in Figure~\ref{fig:phat_figure}.
\begin{equation}
\hat{p}(x) = w(e) =
\begin{cases}
1, & \text{if } e \leq r_f \\
\frac{d(e)}{d(r_f)}, & \text{if } e > r_f
\end{cases}
\label{eq:phat}
\end{equation}

The two core algorithms introduced in \S~\ref{sec:preliminaries} differ subtly in how they apply the given weights, which makes them suitable for different types of applications.
Algorithm A-Chao (\ref{alg:update}), is better suited for sampling from foveated streams because it maintains a weighted accumulation of all previously seen samples, allowing for more temporal continuity. 
When the total weight is high, it indicates that multiple high-quality samples have been likely observed, making it more likely that the current reservoir content reflects a sample with high spatiotemporal fidelity.
In contrast, Algorithm A-ES considers only the weight of the current candidate relative to the sample already in the reservoir, without retaining any information about prior observations. Consequently, it lacks a mechanism for historical integration (for $n < N$) and is prone to discarding valuable high-quality samples too quickly.
We validated this intuition empirically and observed that A-ES indeed tends to lose such samples more frequently than Algorithm~\ref{alg:update}.
We note the observation from \citet{bitterli2020spatiotemporal}, that since our sampling involves temporal accumulation, we can interpret the process as repeatedly combining two reservoirs: one reservoir $R_{1}$, representing the accumulation of all previously seen samples up to the current frame $i$, and another reservoir $R_{2}$ containing just the newly rendered candidate and its associated weight.
As detailed in \S~\ref{sec:wrs_graphics}, combining two reservoirs reduces to treating each as a single weighted sample, while reaping the sampling benefits from the history the reservoir has evaluated.
This observation allows us to use Algorithm~\ref{alg:combine} as proposed in \citet{bitterli2020spatiotemporal}.% , since the key comparison logic in A-ES corresponds exactly to that used in the reservoir combination procedure.
%In our formulation for temporal reuse, $R_{1}$ is the reservoir representing all the past samples until frame $i$, and $R_{2}$ is the reservoir containing the current frame’s rendered sample.

\subsection{Temporal Biasing with Bernoulli Trials}
As previously discussed, Algorithm A-Chao is more appropriate for our application because it accumulates sample weights over time.
However, this same characteristic can become a liability in the presence of a rapidly evolving data stream. 
When the scene changes significantly, the algorithm may continue to give undue importance to previously observed samples that are no longer representative, leading to the rejection of new samples that better reflect the current state.
This is a well-studied problem in reservoir sampling.

Temporal bias functions for reservoir sampling were introduced by \citet{aggarwal2006biased}, which proposes a memory-less bias function to weight reservoir samples lower, such that newer samples have a higher probability to be selected for the reservoir at the time of their evaluation.
This temporally biases reservoir samples based on the relative \textit{time of sampling} compared to when new samples are evaluated.
However, this requires solving for a $\lambda$ that fits the general-case for which the reservoir sampling algorithm is designed, where $\lambda$ is generally very small.
In complex scenes, there is scarcely a constant $\lambda$ that would fit all possible situations.

%Instead, we are interested in ensuring that large scene changes are not discarded by reservoir persistence, but that relevant high-quality samples are not overwritten by poor quality ones too quickly.
Rather than relying on fixed temporal decay, our goal is to ensure that significant scene changes are not ignored due to reservoir inertia, while still preventing low-quality samples from prematurely overwriting important ones that are still perceptually valid.
An alternative solution would be to deterministically preserve the highest weighted sample rendered across frames; however, this strategy assumes that high-fidelity samples will always remain valid or perceptually accurate.
Occlusions / disocclusions, lighting changes, shading rate changes, and filtering effects can all alter the colour of any given pixel; in the case of an occlusion, the scene has fundamentally changed, while for a shading rate reduction, the scene has not changed.
Moreover, diagnosing the specific cause of the reduction can be difficult and expensive, potentially imposing other uncertainty if we were to rely exclusively on the presence of a colour change \cite{mueller2021temporally}.
A deterministic reuse policy would therefore require a robust mechanism to discard samples that have become outdated due to scene or view-dependent changes.
Tuning this mechanism to perform well across the wide range of conditions encountered in immersive environments may incur high complexity and cost; thus, a probabilistic formulation that naturally accounts for this variability and uncertainty could prove more appropriate for high performance situations.

We propose biasing them proportionally to the chance that they survived or not.
As discussed previously, if $\frac{w_{i}}{R_{1} \cdot \wsum + w_{i}} > \rand$, then the current sample held in the reservoir will be replaced; otherwise, it will be maintained in the reservoir.
Therefore, the chance that the sample currently in the reservoir survives is approximately:
\begin{equation}
    \hat{p}(\textrm{survive}) = 1 - \frac{w_{i}}{R_{1_{\wsum}} + w_{i}}
    \label{eq:p_survive_reservoir}
\end{equation}
This model of whether a sample survives the update comparison is a Bernoulli trial, and reservoir samples are repeatedly biased over time using this, with  $w_{i}$ and $w_{sum}$ changing frame-to-frame.
In the case where $R_{1}$ is the previous frame's reservoir (the \textit{temporal} reservoir) and $R_{2}$ is the reservoir for the newly rendered sample (and thus $w_{i} = R_{2_{\wsum}}$), we can bias $R_{1_{\wsum}}$ with the survival rate when it will be compared to $R_{2}$'s sample, giving:

\begin{equation}
    R_{1_{\wsum}} = R_{1_{\wsum}} \cdot \left(1 - \frac{R_{2_{\wsum}}}{R_{1_{\wsum}} + R_{2_{\wsum}}} \right)
    \label{eq:temporal-bias}
\end{equation}

Over k frames, this becomes a Binomial distribution with varying probabilities of survival for every $i < k$ iteration.
After $R_{1}$ is biased according to the incoming probability of the new sample in $R_{2}$, we call the combine function in Algorithm~\ref{alg:combine} to sample the two reservoirs and output a final selected reservoir with an estimate for the best quality pixel sample.

\begin{figure*}[htbp]
    \centering
    \begin{subfigure}[t]{0.32\textwidth}
        \centering
        \includegraphics[width=\textwidth]{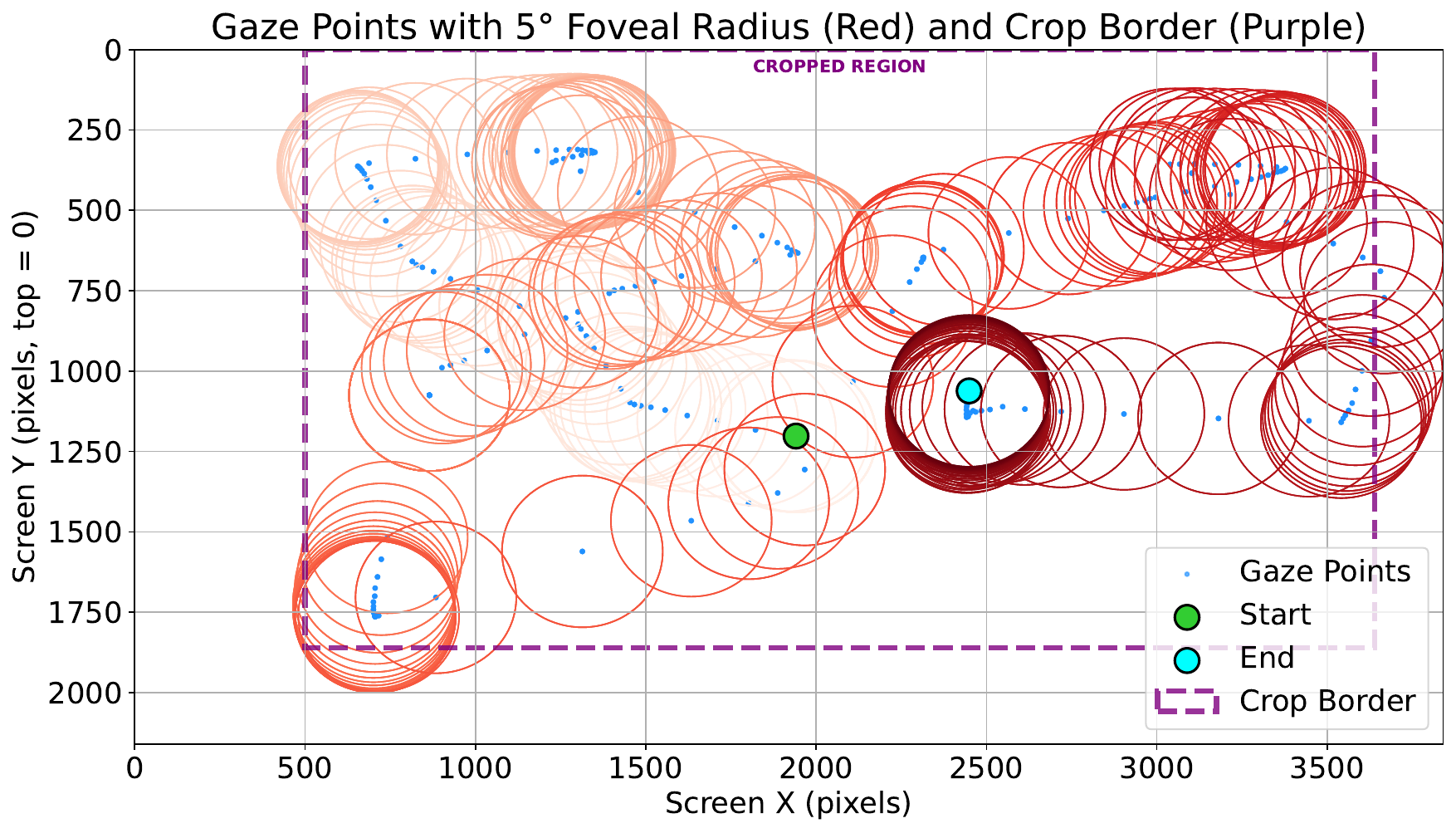}
        \caption{Scanpath and cropping used to generate images in Figure~\ref{fig:vdp_jod}.
        The scanpath started at the center (green point) and moved around the screen to the cyan point.
        Lighter red circles are older points collected in the scanpath.
        Image metrics are evaluated on the full image, but crops are displayed for visual clarity.}
        \label{fig:scanpath}
    \end{subfigure}
    \hfill
    \begin{subfigure}[t]{0.28\textwidth}
        \centering
        \includegraphics[width=\textwidth]{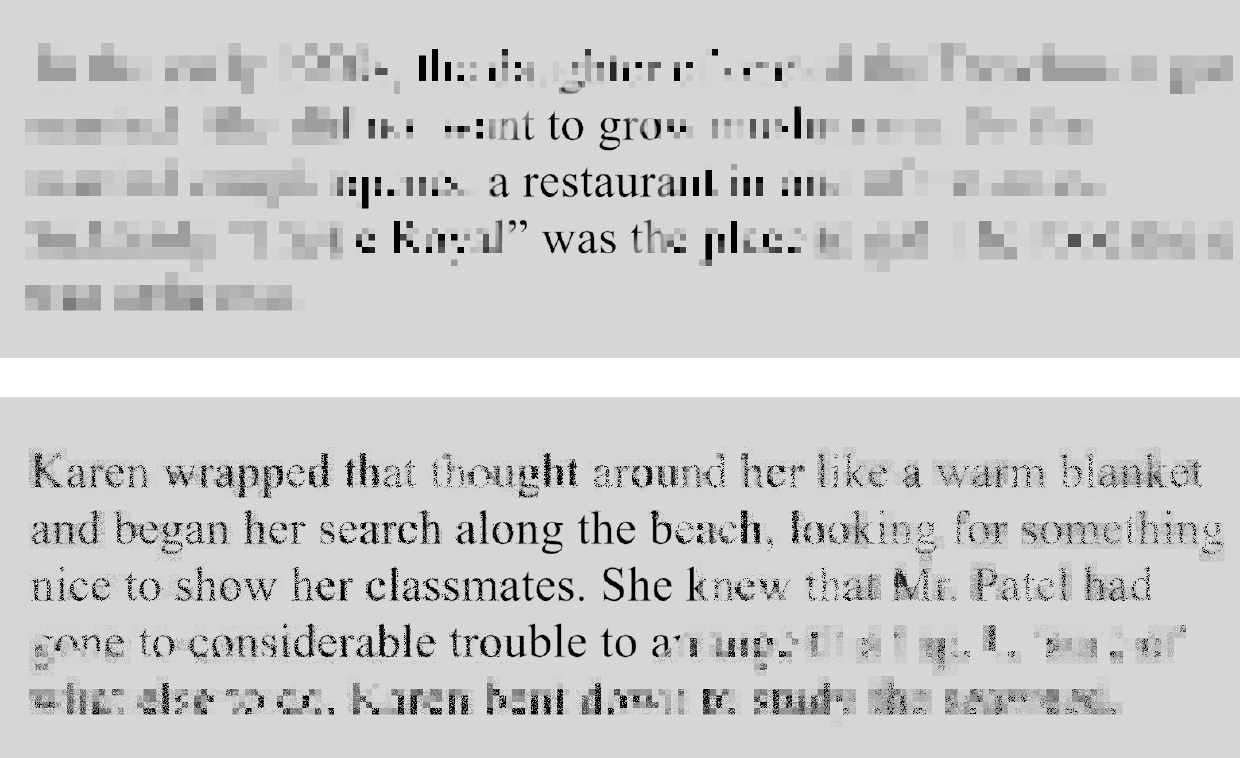}
        %\vspace*{\fill}
        \caption{Example stimuli with a 5\Deg fovea diameter.
        (Top) FOV 5\Deg, (Bottom) WRS 5\Deg.
        Both methods severely limited visual span, but the WRS 5\Deg method temporally preserved samples to be more readable as the eye gaze shifted across the screen during reading.}
        \label{fig:gas_isotherms}
    \end{subfigure}
    \hfill
    \begin{subfigure}[t]{0.34\textwidth}
        \centering
        \includegraphics[width=\textwidth]{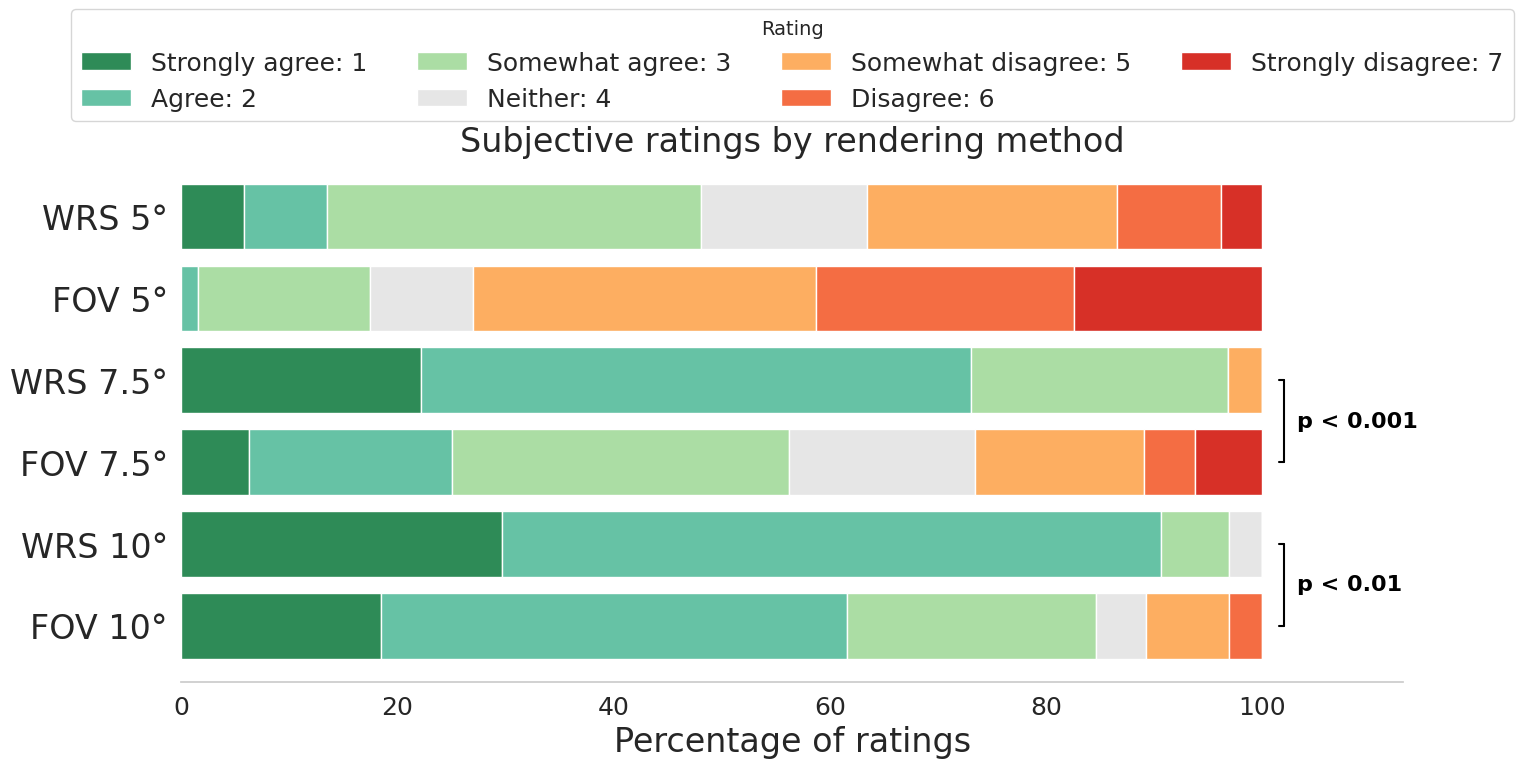}
        \caption{Preference of reading methods for Experiment 1 in \S~\ref{subsec:text_reading}.
        Results show that users significantly preferred WRS 10\Deg vs FOV 10\Deg, and WRS 7.5\Deg vs FOV 7.5\Deg.
        No significant differences in user preference were found between WRS 7.5\Deg vs FOV 10\Deg, WRS 5\Deg vs FOV 10\Deg, and WRS 5\Deg vs FOV 7.5\Deg.}
        \label{fig:barplotwords}
    \end{subfigure}
    %\caption{Combined figure caption (optional)}
    \caption{}
    \label{fig:combined_three}
\end{figure*}

\subsection{Implementation}
We implement our WRS method with temporal biasing using the C\# and HLSL in the Built-In Rendering Pipeline in Unity3D 6.0. 
Foveated rendering is simulated as a postprocess with two methods: The first uses simple mipmapping with bilinear filtering of the output framebuffer,  similar to the shading rate falloffs proposed in \citet{patney2016towards}, and the second uses a progressively increasing Gaussian blur, with blur parameters derived from \citet{tursun2019luminance}.
In raytracing or pathtracing, a lower proportion of peripheral pixels would be accurately estimated, running at a low sampling accuracy per pixel area.
In our rasterizer, shading is performed at a reduced perceptual quality to simulate foveated rendering by computing mipmaps and performing filtering operations.
This is equivalent to rendering foveated pixels into a reduced resolution buffer and then interpolating pixel values.

Temporal reprojection using motion vectors is used to compare temporally neighbouring reservoirs after scene or camera movement has occurred.
Depth buffer comparisons between frames are used to flush a reservoir's temporal sample if an occlusion or disocclusion takes place.
Finally, perceptual colour changes $\Delta L = \texttt{luminance}_{\sample} - \texttt{luminance}_{\sample i}$ are used in our temporal bias function by converting the pixel for the current frame and the temporal neighbour's reservoir sample to LAB colour space (Delta E 94), and then applying the \textit{full} temporal bias function:
% \begin{equation}
%     w_{1_{\textrm{sum}}} = w_{1_{\textrm{sum}}} \cdot (1 - \left| \texttt{luminance}_{\sample1} - \texttt{luminance}_{\sample2} \right|) \cdot (\frac{w_{1_{\textrm{sum}}}}{w_{1_{\textrm{sum}}} + w_{2}})
%     \label{eq:full-temporal-bias}
% \end{equation}
\begin{align}
R_{1_{\wsum}} =  R_{1_{\wsum}}~\cdot~(1 - \left| \Delta L \right|)~\cdot
% \nonumber \\ %
% (1 - \frac{w_{2}} {w_{1_{\textrm{sum}}} + w_{2}})
\left(1 - \frac{R_{2_{\wsum}}}{R_{1_{\wsum}} + R_{2_{\wsum}}} \right)
\label{eq:full-temporal-bias}
\end{align}

\section{Evaluation} \label{sec:evaluation} % todo: change this section name
To evaluate our method, we conducted a series of human subject experiments to understand what impact temporally accumulating foveated samples may have on common user experiences.
The experiments were conducted following a study protocol that was reviewed and approved by our Institutional Review Board (IRB).
We also compare WRS runtime performance and perceptual image metrics (PSNR, SSIM, LPIPS, FovVideoVDP \cite{mantiuk2021fovvideovdp}) against standard foveated rendering, and standard foveated rendering with TAA applied.

\begin{figure*}[h]
    \centering
    \begin{tikzpicture}
        \node[inner sep=0pt] (fig) at (0,0) {
            \begin{minipage}{\textwidth}
                \centering
                % First row
                \begin{subfigure}[b]{0.32\textwidth}
                    \centering
                    \includegraphics[width=\columnwidth]{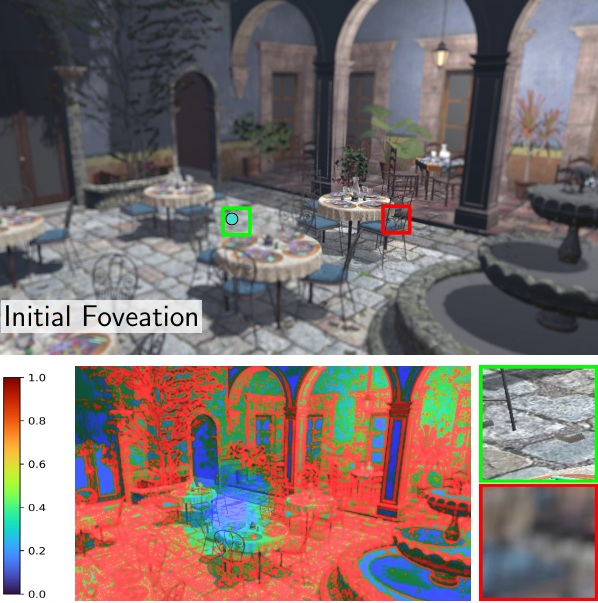}
                \end{subfigure}
                \hfill
                \begin{subfigure}[b]{0.66\textwidth}
                    \centering
                    \includegraphics[width=\columnwidth]{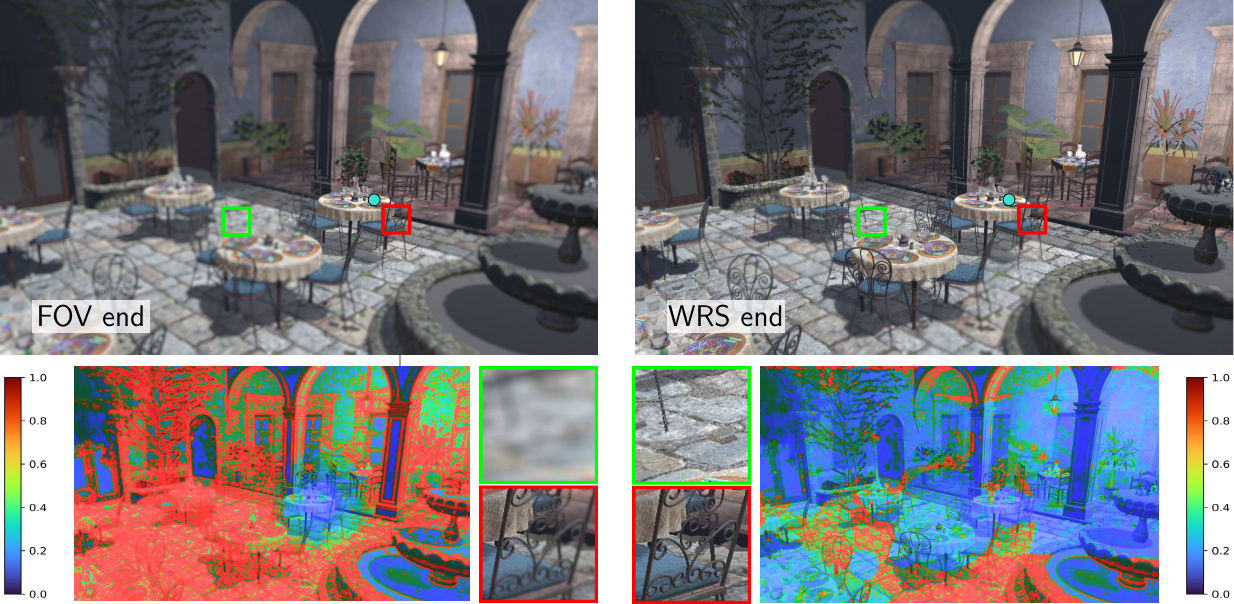}
                \end{subfigure}

                \vspace{1em}

                % Second row
                \begin{subfigure}[b]{0.32\textwidth}
                    \centering
                    \includegraphics[width=\columnwidth]{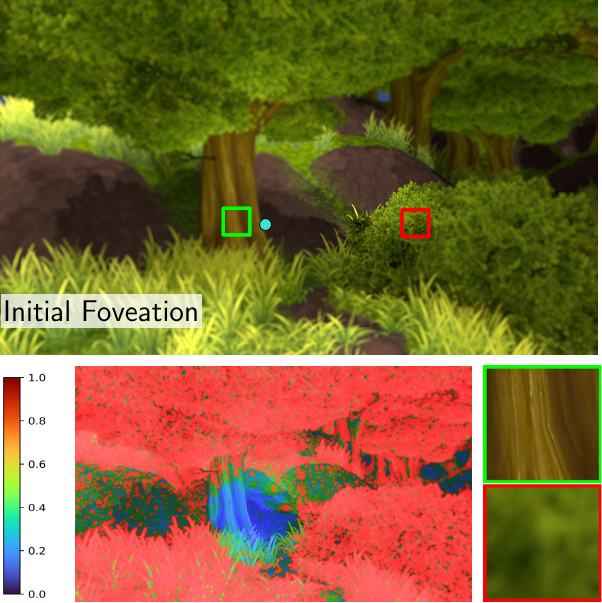}
                    \caption*{Example of FOV JOD Map}
                \end{subfigure}
                \hfill
                \begin{subfigure}[b]{0.66\textwidth}
                    \centering
                    \includegraphics[width=\columnwidth]{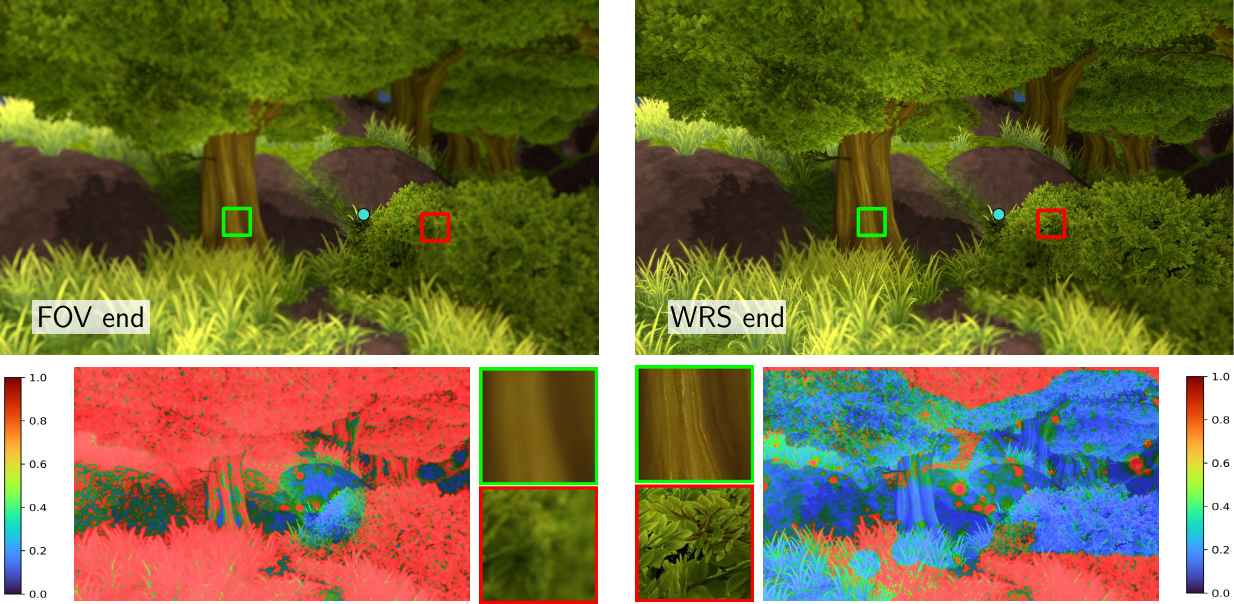}
                    \caption*{Example of WRS JOD Map}
                \end{subfigure}
            \end{minipage}
        };

        % Draw vertical line at center (x=0) across the height of the figure box
        \draw[thick] ([xshift=-0.17\textwidth,yshift=0.4cm]fig.south) -- ++(0, 12cm);

    \end{tikzpicture}

    \caption{
        Images and FovVideoVDP JOD maps for FOV and WRS, both using a 5\Deg fovea, and undergoing the same scanpath as in Figure~\ref{fig:scanpath}.
        Each image started from an initial foveation image on frame 0 (left), and progressed to their final renders (right).
        FovVideoVDP scores are detailed in Table~\ref{tbl:evaluation}.
    }
    \label{fig:vdp_jod}
\end{figure*}

\subsection{Experiment 1: Text Reading} \label{subsec:text_reading}

\begin{figure*}[htbp]
    \centering
    \includegraphics[width=\textwidth]{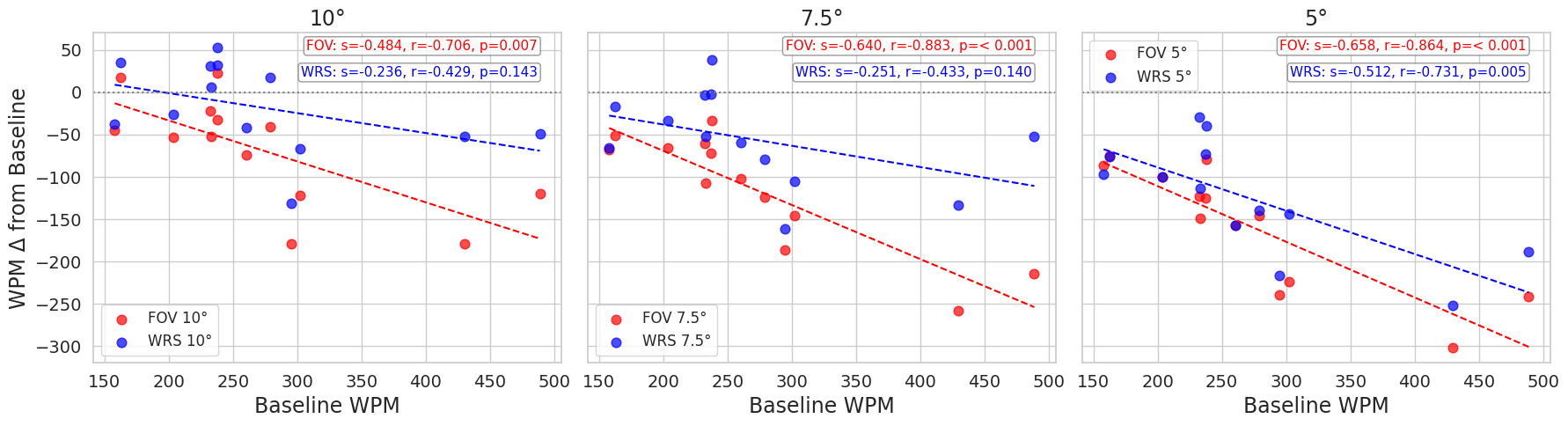}
    \caption{WPM differences from baseline across six rendering conditions.
    For each fovea size, blue shows the WPM decrease under WRS conditions, and red shows the WPM decrease under FOV conditions.
    Slopes (s), Pearson correlations (r), and p values for the decrease are shown.
    WRS 10\Deg and WRS 7.5\Deg did not significantly decrease reading speed, while all other conditions did; in all cases, WRS reduced reading speed significantly less than FOV.}
    \label{fig:wpm_differences}
\end{figure*}

In certain immersive experiences, users may scan their eyes back and forth across the same region repeatedly to understand the image content.
Text reading forces this movement pattern, and thus is an effective way to measure how our sample preservation using WRS can be useful for users.
To evaluate whether our sample accumulation using WRS can help improve reading speed under gaze-contingent foveation, we conducted a text-reading experiment similar to \cite{albert2019reading}.
They found that gaze-contingent foveation can severely impact the reading speed of fast users, but does not have much of an effect on slower readers.
However, their conditions largely do not restrict the visual span of typical readers, and other works that focus on reducing visual span of readers find a decrease in reading speed when visual span is restricted \cite{legge1997psychophysics, chung1998psychophysics, legge2007case}.
We investigate whether our sample accumulation method using WRS can mitigate some negative reading speed reductions due to reduced visual span.

\subsubsection{Experiment Setup}
Fourteen subjects were recruited for the experiment.
One was excluded due to poor performance on the reading comprehension task (< 50\% correct), leaving 13 subjects (age 20-31, M=24.69, 11 male / 2 female).
All participants had normal or corrected-to-normal vision.
Nine subjects were native English speakers, with the remaining all fluent in English.

All stimuli (Figure~\ref{fig:gas_isotherms}) were presented on a 55 inch Samsung LS03D with a 4K (3840x2160 pixels) spatial resolution and 120 Hz temporal resolution.
Peak luminance was $248 cd/m^2$.
Participants' head position was fixed using a chinrest $71.5$ cm from the screen, covering an 80\Deg \space total field of view.
We used the Tobii Eye Tracker 5 running at 133 Hz to track participants' gaze location.
Rendering was performed using Unity3D 6 and an NVIDIA RTX 4070 Ti Super.
All framerates were locked to 120 frames per second.

%\vspace{-5pt}

Users were presented passages from $3^{\textrm{rd}}$ to $5^{\textrm{th}}$ grade reading comprehension exams, ranging from 43 to 57 words.
The stimulus text was presented in the center of the display and spanned $\pm 20\Deg$ horizontal field of view and $10\Deg$  vertical field of view.
The foveal regions showed 20 characters left-to-right at 10\Deg, 15 at 7.5\Deg, and 10 at 5\Deg.
Typical visual span for an English reader is around 15 characters rightward and 3-4 characters leftward \cite{mcconkie1975span}. 
Peripheral nonsense text surrounded the stimuli text and encompassed the remaining peripheral vision.
All text used the Times New Roman serif font.
To severely limit the visual span, we used downsampling with bilinear filtering, rendering every mid-peripheral pixel (from the edge of the foveal region out to 15\Deg diameter = 7.5\Deg radius) at a shading rate of 8x8 pixels, and every far peripheral pixel at a shading rate of 16x16 pixels.
A 1\Deg linear blend was used between regions.

\subsubsection{Procedure}
To begin the experiment, participants were presented three trials to measure baseline reading speed.
Participants were then presented with 6 conditions in random order with 5 unique text passages per condition (30 trials total): Foveated rendering (\textit{FOV}) with foveal sizes of 5\Deg, 7.5\Deg, and 10\Deg, and the same foveal sizes under WRS (\textit{WRS}).
Participants silently read the passages and pressed the keyboard when they were done.
To assess their reading comprehension, they were presented 3 possible word choices that appeared in the paragraph they just read, and asked to select the correct word.
One word appeared in the passage, and the other two were randomly chosen from unique words in other passages.
They then rated how much they agree with the statement, "This text was easy to read under this rendering condition" on a Likert scale from 1 (Strongly Agree) to 7 (Strongly Disagree).
Average accuracy was 77.46\% (SD=8.03\%), and no included participant was below 2 standard deviations from average.

\subsubsection{Results}
WPM differences from baseline were computed for each participant.
Every condition (2 Rendering Condition x 3 Foveal Sizes) passed the Shapiro-Wilks test for normality.
Mauchly's test indicated that the assumption of sphericity had been violated, 
(\( W = 0.026 \), \( \chi^2(14) = 36.97 \), \( p = .001 \)).
Degrees of freedom were corrected using the Greenhouse–Geisser estimate 
\( \varepsilon = 0.412 \).
We use a 2-way repeated-measures ANOVA with the rendering method (FOV vs WRS) and foveal size (5\Deg, 7.5\Deg, 10\Deg) as within-subjects factors.
There was a significant main effect of foveal size on WPM change,
\( F(2, 24) = 108.31 \), \( p < 0.001 \), \( \eta^2_G = 0.302 \),
and a significant main effect of rendering method,
\( F(1, 12) = 31.91 \), \( p < 0.001 \), \( \eta^2_G = 0.128 \).
There was a significant interaction between foveal size and rendering method,
\( F(2, 24) = 4.04 \), \( p = 0.031 \), \( \eta^2_G = 0.008 \).

We found a strong negative correlation between the methodology tested and reading speed in FOV 10\Deg (r = -.71, p = 0.007), FOV 7.5\Deg (r = -.88, p < 0.001), FOV 5\Deg (r = -.86, p < 0.001), and WRS 5\Deg (r = -.73, p = 0.005).
We found moderate negative correlations that were not significant with WRS 10\Deg (r = -0.43, p = .143) and WRS 7.5\Deg (r = -0.43, p = .140).
Full results are shown in \hyperref[fig:wpm_differences]{Figure~\ref*{fig:wpm_differences}}.
While reading speed negatively deviated from baseline as a result of foveation in general across FOV and WRS methods, our results indicate that our temporal sample accumulation using WRS can mitigate some of the negative impacts on reading performance.
Subjective preference scores are shown in \hyperref[fig:barplotwords]{Figure~\ref{fig:barplotwords}}.

\subsection{Experiment 2: Perceptual Validation} 
\label{sec:2ifc} % Moving fixation task ( see S4 in Emulating Foveated Path Tracing https://dl.acm.org/doi/pdf/10.1145/3487983.3488295 and 4.2 in Foveated Real-Time Ray Tracing for Head Mounted Displays)

WRS with sample accumulation was also tested on natural scenes with a moving camera, to ensure that the temporal accumulation under camera movement does not produce temporal artifacts from inaccurate reprojections.
A 2-Interval Forced Choice (2IFC) experiment was run to investigate the utility of our reservoir sampling method (WRS) against standard foveated rendering (FOV) on stimuli that resemble common graphics use-cases more.
The same three fovea sizes as used in Experiment $1$ were used.

\subsubsection{Experiment Setup} \label{subsec:20fcExperimentsetup}
12 participants (age 21-31, M=26.25; 5 female, 7 male) were recruited.
All participants had normal or corrected-to-normal vision.
A similar hardware setup as in \ref{subsec:text_reading} was used.
The screen used was a 27 inch Dell S2725QS with a peak luminance of $262.5 cd/m^2$.
Participants were fixed in a chinrest 42.84 cm from the screen, giving a 70\Deg total field of view.
Because not all the environments used could natively render at 120 frames per second at 4K spatial resolution, the framerate was locked to 60 frames per second.
To keep experiment time low and prevent participant fatigue, participants saw a cascading set of comparisons for the 2IFC task: (FOV 10\Deg \textit{vs} WRS 10\Deg), (FOV 10\Deg \textit{vs} WRS 7.5\Deg), (FOV 10\Deg \textit{vs} WRS 5\Deg), (FOV 7.5\Deg \textit{vs} WRS 7.5\Deg), (FOV 7.5\Deg \textit{vs} WRS 5\Deg), (FOV 5\Deg \textit{vs} WRS 5\Deg).
Users saw 3 scenes (Sponza, San Miguel, Fantasy Forest).

\subsubsection{Procedure}
Participants were presented a cross and asked to fixate on it for the entirety of the stimulus period.
They were presented a moving camera trajectory (4s) that was the same for all conditions.
They would see the first condition (4s), a 0.5s black screen, and then the same camera trajectory (4s) under the second condition.
To maximize the pixel movement, exclusively lateral camera movement was used.
Participants were asked which sequence was more clear.
Because a 2IFC asks the participant to compare the stimulus they most recently saw to their memory of the first stimulus, stimuli were presented in random order such that one condition was shown first 3 times, and the comparison condition was also shown first 3 times. 
In total, users saw 6 comparisons 6 times, across 3 scenes, for 108 trials total.
Trials were discarded and repeated at random if the user's gaze deviated more than 2.5\Deg from the fixation point.
Users were provided breaks between scenes, as well as whenever they desired.
Total experimentation time took around 25 minutes.

%\begin{figure*}[h]
     %\centering%
         %\includegraphics[width=0.90\textwidth]%{figures/images_with_blowups/merged_diagonal_forest.pd%f}
        %\caption{Bottom half of the image: Foveated rendering %following the same scanpath as in %Figure~\ref{fig:vdp_jod}.
        %Top half of the image: The same scanpath but with WRS.
        %In the scanpath, every box shown had at one point been %a foveal sample.}
%        \label{fig:fantasyForest}
%\end{figure*}

\subsubsection{Results} \label{ref:exp3_results}
Since the order of presentation for the reservoir-based and foveated rendering methods was randomized, we conducted a one-sided binomial test for each condition pair to determine whether the observed preference for reservoir rendering exceeded chance. Each condition comprised 216 trials. Participants significantly preferred reservoir rendering over foveated rendering in the following comparisons: WRS~10\Deg{} vs. FOV~10\Deg{} (68\% prefer WRS, \textit{p} < 0.001), 
WRS~7.5\Deg{} vs. FOV~10\Deg{} (57\% prefer WRS, \textit{p} < 0.01),
WRS~7.5\Deg{} vs. FOV~7.5\Deg{} (71\% prefer WRS, \textit{p} < 0.001), 
WRS~5\Deg{} vs. FOV~7.5\Deg{} (62\% prefer WRS, \textit{p} < 0.001), 
and WRS~5\Deg{} vs. FOV~5\Deg{} (64\% prefer WRS, \textit{p} < 0.001). 
Preference for WRS~5\Deg{} vs. FOV~10\Deg{}  did not reach statistical significance (51\% prefer FOV, \textit{p} = 0.076), suggesting that participants could not differentiate between these two methods.
These results show that our method provides better perceived image quality than foveated rendering during camera motion, and can achieve comparable perceptual quality even when rendering a smaller central region.
% Full results are shown in Figure~\ref{subfig:exp3}.

\subsection{Performance}

\subsubsection{Image Quality}
Perceptual image metrics for WRS vs FOV are presented in Table~\ref{tbl:evaluation} for scenes rendered with a 5\Deg foveal region and 15\Deg mid-periphery.
Our WRS method enables high levels of temporal accumulation, allowing for better perceptual image quality.
All sequences were evaluated on PSNR, SSIM, LPIPS, and FovVideoVDP \cite{mantiuk2021fovvideovdp}.
A ~5s scanpath taken from an eye tracker (Figure~\ref{fig:scanpath}) was used to evaluate these metrics.
Images and FovVideoVDP heatmaps for the scanpaths are shown in Figure~\ref{fig:vdp_jod}.
In every case, WRS more closely matches the perceptual quality of the scene without foveation.

To evaluate dynamic camera motion, horizontal back-and-forth motion was used to maximize vection.
We compare WRS to FOV + TAA, and present those results in Table~\ref{tbl:evaluation}.
We find that the temporal reuse of our method outperforms the image quality of TAA applied onto foveated rendering.
In highly reflective scenes, some artifacts appear due to imprecise pixel reprojection with reflections; despite that, WRS reports higher FovVideoVDP scores than FOV + TAA.

\begin{table}[!htbp]
    \centering
    \small
    \setlength{\tabcolsep}{4pt}
    \caption{
        Per-scene perceptual image metrics of WRS and FOV.
        WRS-S and FOV-S use scanpaths generated in Figure~\ref{fig:scanpath}.
        WRS-M and FOV-M use lateral camera motion. FOV-M applies TAA.
        The scanpath was not used on Corridor.
    } 
    \label{tbl:evaluation}
    %\begin{tabular}{llcccccc}
    \begin{tabular}{@{}l@{\hspace{4pt}}l@{\hspace{3pt}}c@{\hspace{3pt}}c@{\hspace{3pt}}c@{\hspace{3pt}}c@{}}
    
        \toprule
        \textbf{Scene} & \textbf{Metric} & \textbf{WRS-S} & \textbf{FOV-S} & \textbf{WRS-M} & \textbf{FOV-M} \\
        \midrule
        \multirow{4}{*}{Fantasy Forest}
          & PSNR $\uparrow$ (dB)      & \textbf{23.71} & 21.93 & \textbf{24.61} & 23.39 \\
          & SSIM $\uparrow$           & \textbf{0.661} & 0.54 & \textbf{0.654} & 0.600 \\
          & LPIPS $\downarrow$        & \textbf{0.178} & 0.409 & \textbf{0.264} & 0.423 \\
          & FovVideoVDP $\uparrow$    & \textbf{7.350} & 6.386 & \textbf{7.717} & 6.676 \\
        \midrule
        \multirow{4}{*}{Sponza}
          & PSNR $\uparrow$ (dB)      & \textbf{26.52} & 24.13 & \textbf{26.52} & 24.10 \\
          & SSIM $\uparrow$           & \textbf{0.720} & 0.540 & \textbf{0.720} & 0.602 \\
          & LPIPS $\downarrow$        & \textbf{0.165} & 0.356 & \textbf{0.165} & 0.388 \\
          & FovVideoVDP $\uparrow$    & \textbf{7.892}  & 7.118 & \textbf{7.892}  & 7.167 \\
        \midrule
        \multirow{4}{*}{San Miguel}
          & PSNR $\uparrow$ (dB)      & \textbf{24.73} & 22.15 & \textbf{21.66} & 20.21 \\
          & SSIM $\uparrow$           & \textbf{0.749} & 0.579 & \textbf{0.693} & 0.614 \\
          & LPIPS $\downarrow$        & \textbf{0.138} & 0.325 & \textbf{0.231} & 0.391 \\
          & FovVideoVDP $\uparrow$    & \textbf{7.928} & 6.993 & \textbf{7.883} & 6.84 \\
        \midrule
        \multirow{4}{*}{Corridor}
          & PSNR $\uparrow$ (dB)      & - & - & \textbf{23.12} & 22.27 \\
          & SSIM $\uparrow$           & - & - & \textbf{0.7752} & 0.7359 \\
          & LPIPS $\downarrow$        & - & - & \textbf{0.2436} & 0.2833 \\
          & FovVideoVDP $\uparrow$    & - & - & \textbf{7.2186} & 7.0636 \\
          
        \bottomrule
    \end{tabular}
\end{table}

\subsubsection{Runtime}
Runtime performance is evaluated on a PC with an Intel i9-12900K CPU (3.2 GHz base), 32GB RAM, and an NVIDIA RTX 4070 Ti S GPU.
We measure the final render performance, from initial rasterization, our foveated postprocess, and finally our WRS method running on top of foveated output.
On all maps, our WRS method runs extremely quickly: 
On Sponza, San Miguel, and Fantasy Forest, all WRS conditions ran in under 0.1ms over foveated rendering while providing much higher perceptual image quality.
On the highly reflective Corridor scene, our method added 0.94ms, but was quicker and higher fidelity than FOV + TAA.

\begin{table}[!htbp]
    \caption{Runtimes (ms/frame) of Weighted Reservoir Sampling (WRS), Foveated Rendering (FOV), and Foveated Rendering with TAA (FOV + TAA).}
    \small
    \setlength{\tabcolsep}{4pt}

    \label{tbl:performance}
    \centering
    \begin{tabular}{lccc}
        \toprule
        \textbf{MAP} & \textbf{WRS} & \textbf{FOV} & \textbf{FOV + TAA} \\
        \midrule
        Fantasy Forest & 12.63 & 12.52 & 14.95 \\
        Sponza         & 9.82  & 9.77  & 11.97 \\
        San Miguel     & 13.59 & 13.53 & 18.62 \\
        %Coffee Shop    & 37.52 & 37.36 & 38.59 \\
        Corridor       & 10.81 & 9.87 & 11.66 \\
        \bottomrule
    \end{tabular}
\end{table}

% \begin{tabular}{|l|c|c|}
% \hline
% \textbf{Metric} & \textbf{My Method} & \textbf{Baseline} \\ \hline
% LPIPS $\downarrow$ & 0.12 & 0.15 \\ \hline
% PSNR $\uparrow$ (dB) & 28.5 & 26.2 \\ \hline
% SSIM $\uparrow$ & 0.95 & 0.92 \\ \hline
% FovVideoVDP $\uparrow$ (JOD) & 6.8 & 7.2 \\ \hline
% \end{tabular}

\section{Limitations}
\begin{figure}
    \centering
    \includegraphics[width=\columnwidth]{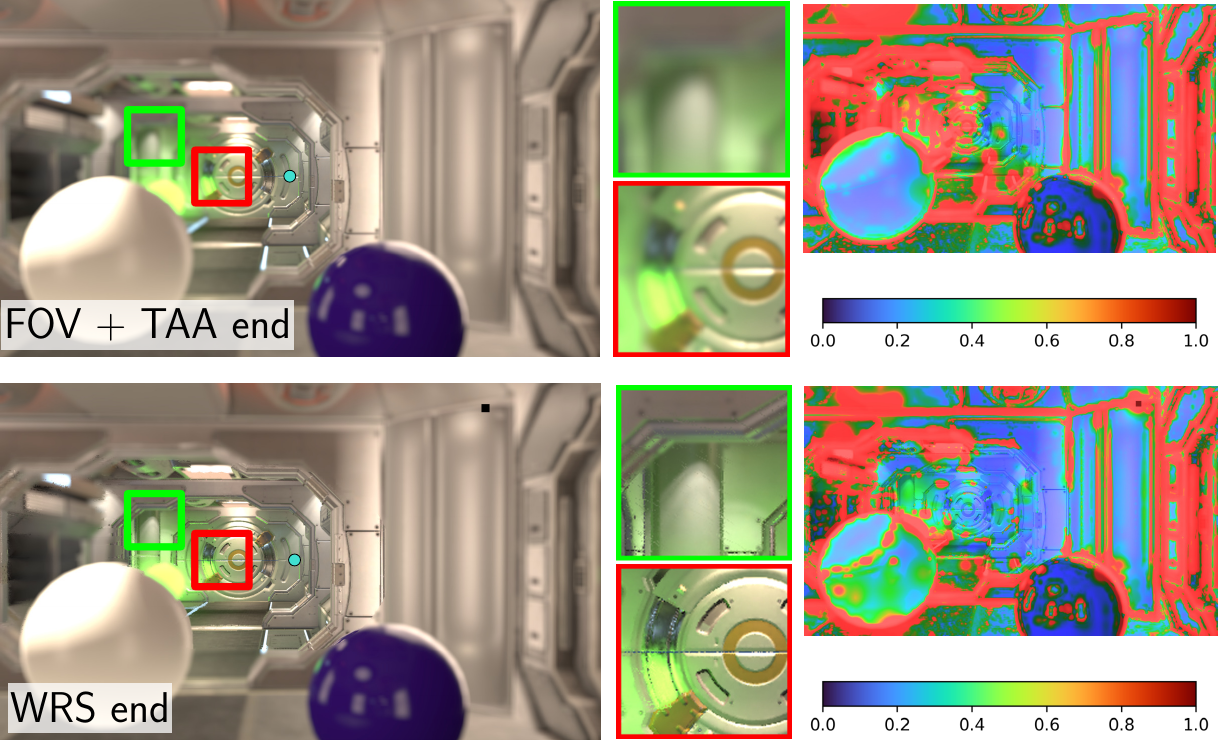}
    \caption{
        Highly reflective Corridor scene with dynamic objects and camera motion and rotation, rendered with FOV + TAA (top) and WRS (bottom).
        In rasterization-based reflective renders, simply using motion vectors can lead to inaccurate reprojection, causing artifacts to appear when temporally accumulating samples, as pixels  get reprojected to incorrect locations, impacting the reflection quality.
        Despite this, WRS had higher FovVideoVDP scores.
    }

    \label{fig:corridor}

\end{figure}

Our results show positive improvements over traditional foveated rendering, and foveated rendering with other temporal reuse methods applied.
However, our method is imperfect; when object occlusion or disocclusion occurs, we fall back to strict foveated rendering, forcing a reset of the temporal history of the reservoir.
In scenes with many dynamic objects, this limits the efficacy of our implementation.
Additionally, as our method works on foveated rasterization, reflections and refractions can be difficult to reproject properly, as the motion vectors are no longer accurate.
This can lead to artifacting from incorrect reprojection, producing a speckled effect, seen on the highly reflective Corridor scene (Figure~\ref{fig:corridor}).
Despite this, our WRS method produces higher FovVideoVDP quality over FOV + TAA, which is promising for future improvements.
%Full results are available in Table~\ref{tbl:evaluation}.
% Table 1 already referenced in 5.3.1

\section{Conclusion and Discussion}
Foveated rendering allows for computational savings for high-resolution rendering tasks by rendering peripheral regions at a lower spatial resolution.
Traditional methods, however, do not tend to temporally save previously high-resolution pixels between renders, potentially losing out on opportunities due to saccadic eye behaviour.
In this paper, we introduced Weighted Reservoir Sampling to Foveated Rendering in rasterization contexts to temporally accumulate high-quality samples.
Our key insight is that during fixation, saccades, and upon saccadic landing, smaller ocular drifts and microsaccades occur to enhance details in foveal vision.
With traditional eye-tracked foveated rendering approaches, any samples that were foveal samples in a prior frame but not in the current one will be discarded and rendered at a lower quality resolution.
However, by stochastically accumulating perceptually relevant frames and combining that with a new temporal bias scheme based on a Binomial distribution, we can achieve high levels of temporal accumulation at a very low computational cost.

\bibliographystyle{ACM-Reference-Format}
\bibliography{ville-bib}

\end{document}